\documentclass[runningheads]{llncs}
\usepackage{graphicx}
\usepackage{csquotes}
\usepackage{paralist}
\let\oldFootnote\footnote
\newcommand\nextToken\relax

\renewcommand\footnote[1]{%
    \oldFootnote{#1}\futurelet\nextToken\isFootnote}

\newcommand\isFootnote{%
    \ifx\footnote\nextToken\textsuperscript{,}\fi}

\usepackage[breaklinks]{hyperref}
\expandafter\def\expandafter\UrlBreaks\expandafter{\UrlBreaks
  \do\a\do\b\do\c\do\d\do\e\do\f\do\g\do\h\do\i\do\j%
  \do\k\do\l\do\m\do\n\do\o\do\p\do\q\do\r\do\s\do\t%
  \do\u\do\v\do\w\do\x\do\y\do\z\do\A\do\B\do\C\do\D%
  \do\E\do\F\do\G\do\H\do\I\do\J\do\K\do\L\do\M\do\N%
  \do\O\do\P\do\Q\do\R\do\S\do\T\do\U\do\V\do\W\do\X%
  \do\Y\do\Z}
\usepackage[nolist]{acronym}
\usepackage{listings}
\usepackage{booktabs}
\usepackage{tabularx}

\usepackage{silence}
\usepackage{caption}
\usepackage{subcaption}

\newcommand{\ph}{Provenance Holder}

\begin{document}
\title{Trusted Provenance of Automated, Collaborative and Adaptive Data Processing Pipelines}
\titlerunning{Trusted Provenance of Data Processing Pipelines}
%
\author{Ludwig Stage \and Dimka Karastoyanova}
\authorrunning{L. Stage, D. Karastoyanova}
%
\institute{Information Systems Group, University of Groningen, The Netherlands\\
\email{bpm@ludwig-stage.de}, \email{d.karastoyanova@rug.nl}}
\maketitle              
\begin{abstract}
To benefit from the abundance of data and the insights it brings \textit{data processing pipelines} are being used in many areas of research and development in both industry and academia.
One approach to automating data processing pipelines is the workflow technology, as it also supports collaborative, trial-and-error experimentation with the pipeline architecture in different application domains. In addition to the necessary flexibility that such pipelines need to possess, in collaborative settings cross-organisational interactions are plagued by lack of trust. While capturing provenance information related to the pipeline execution and the processed data is a first step towards enabling trusted collaborations, the current solutions do not allow for provenance of the change in the processing pipelines, where the subject of change can be made on any aspect of the workflow implementing the pipeline and on the data used while the pipeline is being executed. Therefore in this work we provide a solution architecture and a proof of concept implementation of a service, called \ph{}, which enable \textit{provenance of collaborative, adaptive data processing pipelines in a trusted manner}. We also contribute a definition of a set of properties of such a service and identify future research directions.

\end{abstract}

\begin{acronym}
	\acro{ACID}{Atomicity, Consistency, Isolation, Durability}
	\acro{BPEL}{Business Process Execution Language}
	\acro{BPM}{Business Process Management}
	\acro{BPMS}{Business Process Management System}
	\acro{EPR}{Endpoint Reference}
	\acro{ESB}{Enterprise Service Bus}
	\acro{FAIR}{Findable Accessible Interoperable Reusable}
	\acro{PoC}{Proof of Concept}
	\acro{SLA}{Service-level agreement}
	\acro{SOA}{Service-oriented Architecture}
	\acro{SOC}{Service-oriented Computing}
	\acro{RARE}{Robust Accountable Reproducible Explained}
   \acro{UDDI}{Universal Description, Discovery and Integration}
   \acro{UUID}{Universally Unique Identifier}
   \acro{WSDL}{Web Services Description Language}
   \acro{WfMS}{Workflow Management System}
   \acro{sWfMS}{Scientific Workflow Management System}
\end{acronym}
 
\section{Introduction}
\label{sec:introduction}

A significant part of data-driven ICT research and development in enterprises heavily relies on data analysis, simulations and machine learning algorithms. Recently, in a wave of initiatives towards supporting wide-spread transformation to a digital world, there has been an enormous effort by both enterprises in different industries and research to automate and deploy data processing onto the enterprise computing environments in order improve their operations and to benefit the most from the available data. 

The first step towards this goal is the automation of the computational and data processing steps needed using data processing pipelines that can be implemented in many different ways using different methodologies and technologies. The major challenges such a task faces are related to identifying what the best approach is towards the actual automation of the data pipeline and integration of the computational resources, the ability to use data from different sources in different formats and varying quality properties, the flexibility of the data pipelines, the modularity and reusability of individual steps, the ability to enable collaborative modelling and execution of data processing pipelines, as well as their provenance and reproducibility. All these challenges have been in the focus of research and industries for quite some time and there is abundant literature reporting on interdisciplinary research results from many different communities, like data science, intelligent systems, scientific computing and workflows, eScience and others, employing a huge variety of concepts and technologies. 

The topic of \textit{provenance}\footnote{"The provenance of digital objects represents their origins." source: https://www.w3.org/TR/2013/NOTE-prov-primer-20130430/} has been researched predominantly in the field of scientific experiments and scientific workflows, which led to the definition of the characteristics of \ac{FAIR} results \cite{Mesirov415,FAIR} and \ac{RARE} experiments \cite{CaroleGoble15}. In this field, scientific experiments are considered to be of good provenance if they are reproducible \cite{ABJF06}. Enabling reproducibility of experiment results, typically by means of tracking the data through all processing, analysis and interpretation steps of the experiment, has been one of the main objectives of scientific workflow systems, in addition to  the actual automation of scientific experiments. The importance of provenance in in-silico experiments has been identified, discussed and approaches have been partly implemented more recently in e.g. \cite{atkinson:hal-01544818,herschel2017survey,TavernaDataProvenance,Taverna13,DBLP:journals/debu/FreireC17} and are relevant to enabling the provenance of data processing pipelines.
Furthermore, there are initiatives towards standardization of representing provenance information for the purposes of both modeling provenance information and establishing an interchangeable format for such information such as PROV-DM\footnote{https://www.w3.org/TR/prov-overview/}.

The scope of our work includes automated data processing pipelines, which  use only software implementations of computational and data transformation tasks and excludes data processing pipelines in which participation of physical devices (e.g. microscopes, wet labs, sensors and actuators) is directly visible in the pipeline. Having said that, we focus additionally on enabling the \textit{provenance of flexible, a.k.a. adaptive, data processing pipelines that are carried out in collaboration} among identifiable organisational entities. The matter of \textit{trust among the collaborating parties} is of utmost importance in the context of our work, in particular because of the need to capture the origins of change that can be carried out by any of the participating parties at any point in the execution of the pipelines.

Our technology of choice for modelling and running collaborative data processing pipelines is \textit{service-based, adaptable processes, both workflows and choreographies}, that are well known from the field of \ac{BPM} \cite{Weske19} and conventional Workflow Management Technology \cite{Leymann2000_ProductionWorkflow} for their beneficial properties such as modularity, reusability, interpretability, transactional support, scalability and reliability. In other related research of ours we have provided a \ac{WfMS} supporting the execution of adaptable, collaborative choreographies and workflows and have also evaluated its applicability in the domain of scientific workflow automation \cite{Model-as-you-go}.

To the best of our knowledge, the ability to reproduce the changes on either workflow or choreography models or instances made by collaborating organisations in the course of running their data processing pipelines in a trusted manner, has not been the subject of other works. We call this type of of provenance \enquote{\textit{trusted provenance of change}}.

Towards closing this gap in research, we extend our vision of a solution\cite{BPM2019}, called \textit{\ph{} service}, that has to track and record all changes made on choreography and/or workflow models or instances to support their provenance in a trusted manner and allow collaborating organisations to retrace and reproduce their data processing pipelines exactly the same way as they have been carried out, including all changes made on both data and software used during the execution. The contributions of this work are: (i) An extension of the workflow provenance taxonomy to account for adaptation (ii)  Detailed definition of the properties of the \ph{} service that will guarantee trusted provenance of collaborative, adaptive data processing pipelines, (iii) functional architecture, which is generic in nature and applicable in any application domain and imposes low effort to integrate with other flexible \ac{WfMS} systems and (iv) an implementation as a proof of concept. 
We also explicitly identify (v) the \textit{prerequisites} for employing the \ph{} with other \ac{WfMS} environments, namely the ability to support the trial-and-error manner of experimenting (as in e.g. Model-as-you-go-approach \cite{ART-2013-06} or ability to change and propagate change in choreographies \cite{Rinderle-Ma_ChorChangePropagation2015}) and the ability to provide workflow monitoring data that allows for data and workflow provenance \cite{BPM2019}.

The paper structure is as follows. In \autoref{sec:requirements-provenancetypes-properties} we reiterate the requirements on our system, illustrate the supported provenance types and define the properties of a system that can enable provenance and trust in adaptive collaborative data processing pipelines, while the architecture of the \ph{} supporting these properties is described in depth in  \autoref{sec:architecture}.
In \autoref{sec:implementation} we contribute a \ac{PoC} implementation of the \ph{} and elaborate on different design and implementation details and discuss our design decisions.
In \autoref{sec:discussion} we identify open issues and directions for future research and  \autoref{sec:conclusions} concludes the paper.

\section{\ph{}: Requirements, Supported Types of Provenance and Properties}
\label{sec:requirements-provenancetypes-properties}

In this section we reiterate the requirements for a system enabling reproducible, trusted and adaptive collaborations, we discuss existing provenance types and expand these with new types of provenance, as well as we present the \ph{} properties.

\subsection{Requirements}
\label{sec:PHrequirements}

Previously we identified four requirements  in \cite{CAiSE18} and \cite{BPM2019} for reproducible, trusted and adaptive collaborations, e.g. scientific experiments (including eScience).
Furthermore we argued these requirements are to be fulfilled by an enabling system.
Hence, these requirements are to be provided by the \ph{}.
We numbered these requirements (cf. \autoref{tab:requirements}) for a better overview, the possibility for clear referencing and mapping purposes. 

\setlength{\tabcolsep}{4pt}
\begin{table}[htp]
\centering
		\caption{\ph{} Requirements adopted from \cite{CAiSE18,BPM2019}}
		\label{tab:requirements}
\begin{tabular}{c l}
	\toprule
	\textbf{Requirement}  &  \textbf{Description} \\
	\midrule
	\textbf{R1} & \textbf{Adaptability}  to adhere to the adaptability of experiments \\
	\textbf{R2} & \textbf{Provenance} to enable \ac{FAIR} results \cite{Mesirov415} \\
	\textbf{R3} & \textbf{Reproducibility}  for \ac{RARE} experiments \cite{CaroleGoble15} \\
	\textbf{R4} & \textbf{Trust} among collaborating parties to also enable accountability \\
	\bottomrule
\end{tabular}
\end{table}

\subsection{Supported Types of Provenance}
\label{sec:PHprovanancetypes}

According to the most recent survey on provenance \cite{herschel2017survey}, there are four types of provenance: provenance meta-data, information system provenance, workflow provenance and data provenance; in all cases artefacts are considered to be of good provenance if their origin and history of transformations have been sufficiently recorded to be reproducible. While provenance meta-data is the least specific one,  data provenance is considered  the most specific one and is also known as data lineage.

\begin{figure}
    \centering
    \includegraphics[width=\textwidth]{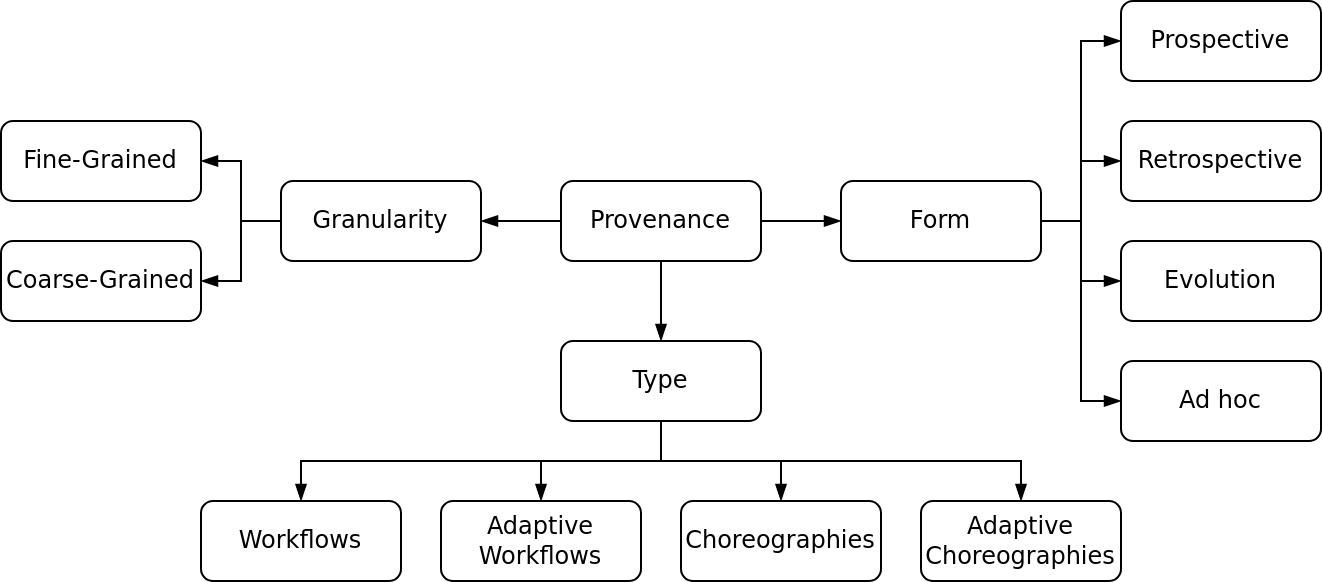}
    \caption{Workflow Provenance types taxonomy}
    \label{fig:provenance-types}
\end{figure}

The workflow provenance type (see Figure~\ref{fig:provenance-types}) directly applies to our use case of adaptive data processing pipelines, hence it is the type we focus on in this work.
In addition to the provenance information regarding the control and data flow of workflows, workflow provenance  includes input, output and parameters of the workflows; collecting such provenance information requires appropriate instrumentation of the \ac{WfMS}. Furthermore, the authors of \cite{herschel2017survey} group workflow provenance in form (prospective, retrospective, evolution) and granularity (coarse-grained and fine-grained).
Prospective provenance  captures only workflow models and their (run-time) contexts, retrospective provenance captures  additionally the input data, whereas evolution provenance captures the changes on models, input data or context.
In addition to that classification of workflow provenance, there is a need to account for the different types of \textit{provenance of adaptation or change} of workflows, as this is not part of the work of \cite{herschel2017survey} but is necessary for enabling provenance of adaptive workflows (or trusted provenance of change). We therefore subdivide provenance of adaptation/change  into \textit{workflow evolution provenance} and \textit{provenance of ad-hoc workflow change}.
This distinction is important, as the subject of the change are either workflow model or one or more workflow instances, respectively, and ensuring their provenance is addressed using different approaches and require different data.
The former type of change is typically enacted using instance migration from one workflow model to another, whereas the latter is carried out directly on the internal representation of a process instance running on a workflow engine.
Furthermore, there is a need to distinguish between the provenance forms \textit{provenance of adaptive workflows} and \textit{provenance of adaptive choreographies}, as in collaborative data processing pipelines the changes in a choreography (model or instances) of workflows have to be tracked and captured, too.

The ability to support these types of provenance implies specific requirements on the \ph{} service. Namely, it has to be able to track the models and model adaptation/changes, instance migrations and ad- hoc changes, of both workflows and choreographies, as well as the executions of choreographies and workflows which produced a certain output given a certain input (data and other parameters).
Since we consider a collaborative environment, traces of workflow execution or even actual changes to models are not (immediately) published for confidentiality and trust reasons. For the \ph{} service it means that it will have to capture only representations of the actual objects containing the relevant information rather than the actual detailed workflow execution and data traces. In the terms of \cite{herschel2017survey} this means that the \ph{} records (representations of) models (W), input data (D) and the run-time context (C) and changes thereof. The types of provenance to be supported are the prospective, evolution and coarse-grained ones. Depending on the level of detail of available execution traces, retrospective and even fine-grained provenance can be enabled, too. The \ph{} is required to support also the provenance of adaptation/change in form of workflow evolution provenance and provenance of ad-hoc workflow change (as per Figure~\ref{fig:provenance-types}).

In the next section we focus on the properties a \ph{} must possess in order to support all these provenance types and meet the requirements R1 through R4 (Table \ref{tab:requirements}).

\subsection{\ph{} Properties}
\label{sec:PHproperties}

Note that in our work we assume only a minimum level of trust. This accounts for the fact that process and/or choreography participants may prefer to keep the details about their data processing pipeline confidential until its potential disclosure (or forever). At the same time we aim at giving a choreography participant the possibility to make statements about their data processing steps that can be trusted by other parties. That is why a \ph{} service is required to keep the provenance information separately from the data and execution of the data processing pipeline and thus does not disclose publicly any insights about the actual data processing.

To enable the provenance of collaborative adaptive choreographies and meet the above explained requirements, the \ph{}{} has to enable 
a choreography participant (referred to as \textbf{I} in Table~\ref{tab:properties}) to make the following four statements about their data processing pipelines without directly disclosing inner workings or data. A choreography participant can be person or system invoking a workflow/choreography, deploying/adapting a workflow/choreography model or executing such a workflow/choreography (hence a participant is using an appropriate system for that, e.g. \ac{WfMS}).
We map these statements to \ph{} properties, as they imply guarantees for specific capabilities of the service (cf. \autoref{tab:properties}).

\renewcommand{\arraystretch}{1.2}
\begin{table}[h!t]
\centering
		\caption{\ph{} Properties and their mapping to statements made by choreography participants. In the statement column the pronoun \textbf{It} is information about either of the following:  result, origin/predecessor or change. The text in bold highlights where the focus of each property lies.}
		\label{tab:properties}
\begin{tabularx}{\textwidth}{c c X}
	\toprule
	\textbf{Property} & \textbf{Statement by participant} &  \textbf{Description} \\
	\midrule
	
	\textbf{P1} & \enquote{\textbf{I} know it} & A result/change/predecessor can be attributed to a certain identifiable entity, i.e. choreography participant. \\
	
	
	\textbf{P2} & \enquote{I knew it \textbf{before}} & A result/change/predecessor has been available/known or has happened at or before a certain point in time. \\
	
	
	\textbf{P3} & \enquote{I \textbf{actually} know it} & Prove that that participants know of a result/change/predecessor (without information disclosure). \\
	
	
	\textbf{P4} & \enquote{I know \textbf{where it came from}} & Participants have knowledge of the predecessor of a result/change/predecessor. \\
	\bottomrule
\end{tabularx}
\end{table}

While in this section we highlight some of the concepts and techniques that can be used to enable these four properties, we will give more details later in this article. 

To be able to attribute something to someone (P1) we employ public/private key digital signature because of its widespread use and because after an initial identification or pairing of a party to public key authenticity can easily be verified.

With time stamping, e.g. on an immutable public ledger (like e.g. blockchain), it will be possible to prove that
something was known to have happened at a certain point in time (P2) or at least that it was known before a certain point in time. Time stamping via blockchain proves that something was known before; alternatively, the signature of an execution timestamp by invoker and executor might be considered but is not as strong of an indicator.

Proving that someone knows a particular something can  be trivially and obviously achieved by simply disclosing said information.
Therefore, property P3 can always be achieved through information disclosure.
However, in order to prove that someone knows a particular something without disclosing sensitive information (P3), we will investigate further the concept of zero knowledge proofs (ZKP) and more specifically non-interactive zero-knowledge proofs (\cite{tariq2023verifiable}, as it presents a systematic overview over the greater topic of verifiable privacy-preserving computations).
Examples of application for non-interactive zero-knowledge proofs include the Zerocash protocol 
of the cryptocurrency Zcash\footnote{\url{https://github.com/zcash/zcash}}\footnote{\url{https://z.cash/}} and the Monero cryptocurrency\footnote{\url{https://www.getmonero.org/}}.

Property P4 requires the three previous properties and is meant to prove the origin of something by linking it to its predecessor. This can be done by combining the properties P1, P2 and P3.

While the support of properties P1 and P4 by the \ph{}, together with the actual data, models and changes thereof, guarantees the aforementioned workflow provenance types, the properties P2 and P3 address the issue of ensuring trust in adaptive collaborations.
A third party can view the recorded provenance information and even verify properties P1, P2, (P3) and P4 without engaging in collaboration or with a respective choreography participant.
In this work we will concentrate on the design and implementation of three out of these four properties, more precisely on P1, P2 and P4 (see \autoref{ssec:methods-operations}), whereas P3 will be subject of our future research.
In the following section we present the detailed architecture of the \ph{} with focus on the properties presented here. 

\section{\ph{} Architecture}
\label{sec:architecture}

The \ph{} is a service responsible for collecting all information necessary to ensure provenance and reproducibility of and trust in the collaborative adaptations and enable the four properties we introduced in the previous section (\autoref{sec:PHproperties}). We aim at providing a generic, reusable and non-intrusive solution across different scenarios and separation of concerns \cite{Dijkstra1982}.

\begin{figure}[htp]
    \centering
    \includegraphics[width=.8\textwidth]{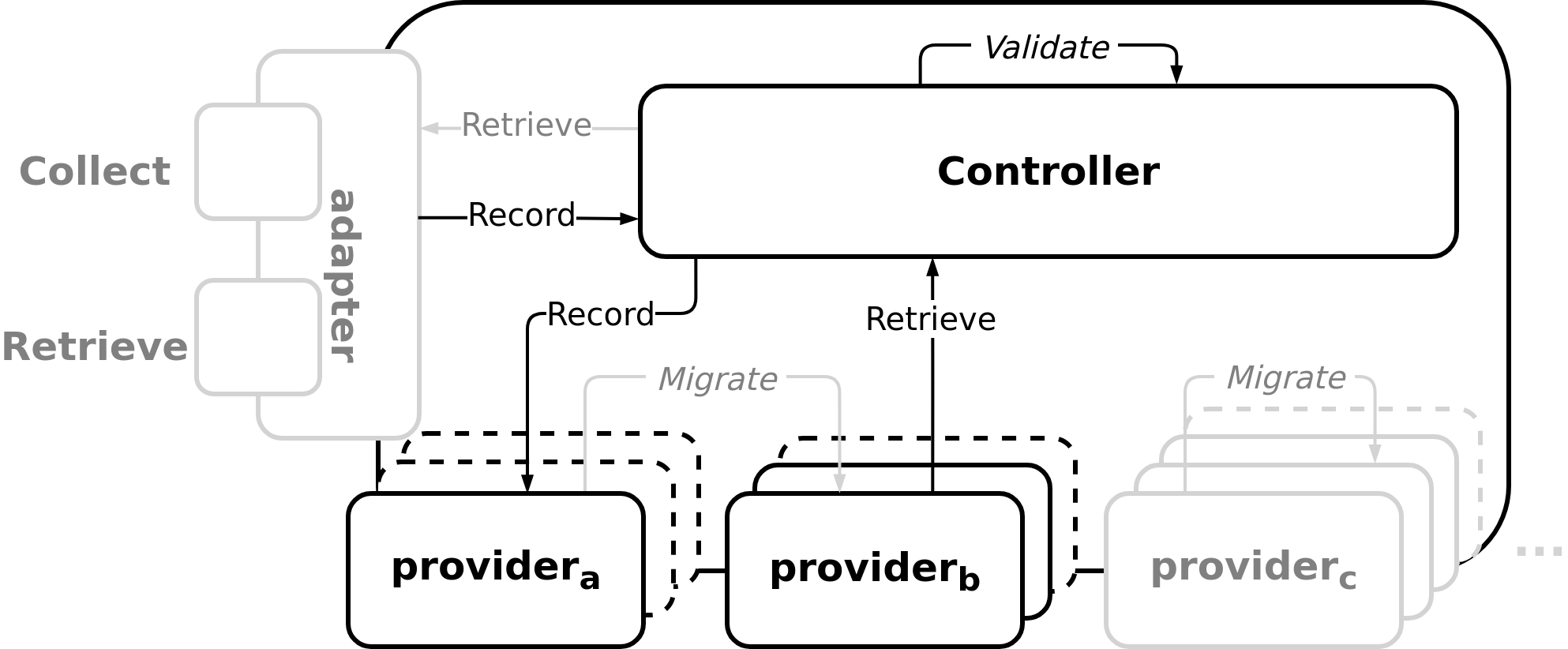}
    \caption{\ph{} Architecture: components, external operations and internal methods, implemented ones are black (adopted from \cite{BPM2019})}
    \label{fig:arch}
\end{figure}

The \ph{} service provides \textit{two main operations} as part of its interface (cf. \autoref{fig:arch}): 1) \textit{collect provenance data} (\textit{Collect}) and 2) \textit{retrieve provenance information} (\textit{Retrieve}); we call these operations also  external operations.
The controller, the adapter and one or more provenance providers are the \textit{components of the \ph{}} (cf.  \autoref{ssec:components} and  \autoref{fig:arch}) and they carry out four \textit{interaction scenarios} in order to  realize the two externally provided operations of the \ph{} service
(see \autoref{ssec:methods-operations}). The interaction scenarios are always combinations of several of the internal methods\footnote{We use the term \textit{method} for disambiguation purposes only.}; the (internal) methods are: \textit{Record}, \textit{Retrieve}, \textit{Validate} and \textit{Migrate}. 

\textit{Collection of provenance data} is done through selection of the relevant messages from the service middleware used for the interaction among participants in the choreography (e.g. and \ac{ESB} or  any other data transfer technology).
We regard every piece of data related to execution of workflows and choreographies or their adaptation that is communicated via the service middleware  as provenance data. After selection, processing and storage by the \ph{} it becomes \textit{provenance information}. In this process, the collected provenance data is validated by the Validate method provided by the Controller component and recorded using the Record operation of one of the Provider components.

\textit{Retrieving provenance information}, i.e. \textit{Retrieve} operation, is done by using the methods Retrieve and Validate.
Afterwards the provenance information is published to the service middleware so that a Provenance viewing or visualization tool can be used by an expert to inspect the provenance information.

\subsection{Components}
\label{ssec:components}

As shown in \autoref{fig:arch}, the components of the \ph{} are the controller, the adapter and one or more provider components.

The \textit{adapter} is the component ensuring the \textit{integration} of the \ph{} with other systems. Its actual design and implementation are specific for the system with which it has to work to enable the integration and correct communication. We recommend using a service middleware that facilitates the integration of workflow management systems and service-based components with other functionality, however we do not assume any specific technology in our generic solution.  
In addition it provides the two external operations: \textit{Collect} and \textit{Retrieve}.
To do so, the adapter acts as interface to external existing system and has to be directly connected to the communication middleware which transports the provenance data.
Furthermore, the adapter has the important task to identify the appropriate data to pick from the service middleware and hand it over to the controller component for further processing. The actual interaction with the service middleware has to comply with the adapter interface, i.e. use the external operations, but the implementation has to be dealt with individually for each software landscape.
We envision two possible approaches for the \textit{selection of provenance data}: i) it can be actively provided by the \acp{WfMS} running the choreographies and the middleware they use or ii) the adapter component has to carry out the selection of relevant data.
Both approaches have their own advantages and disadvantages and are going to be discussed in \autoref{sec:discussion}.

In terms of \textit{adaptations} made on workflows and choreographies we distinguish between changes which cause an instance migration and those that stand for ad-hoc changes.
Capturing the former is done by recording the new model. The latter case is more difficult to tackle because changes are applied to a model but do not necessarily generate a new model representation to keep track of.
Furthermore, changes can be applied consecutively to a \enquote{base} model and they might add to it or remove previously added parts.
Changes might evolve to a new model version or they might be abandoned altogether.
While capturing these changes and keeping track of them is not trivial, it also becomes apparent that keeping track of these transient changes is important, too, 
since they are not inherently captured through manifestation in a model version.
Distinguishing between execution and adaption data, as well as differentiating between instance migration and ad-hoc change is a task of the adapter component.

\textit{Data retrieval}, which scientists commonly call publication, is done only upon request sent by the user typically via a service or tool capable of presenting the provenance information - both the actual information (if provided by the involved participants) and the fact that it can be trusted.
The adapter component serves the request with only the collected provenance information, though.

To enable both provenance data collection and retrieval imposes additional requirements, especially on the \ac{WfMS} and software used to model and monitor the workflows and choreographies (see \autoref{sec:requirements-on-wfms}).

\textit{Providers} or \textit{provenance providers}
have to implement three out of the four methods explained in \autoref{ssec:methods-operations}, namely \textit{record, retrieve and migrate} and certain requirements to fulfil.
The implementation complexity can be arbitrary high and strongly depends on the employed technology: writing into a log file, for example, is of low complexity, whereas employing blockchain technology is more on the high end of the complexity spectrum. The needs of different workflow types also come into play when deciding which technology to use.

The \textit{controller}
is in charge of the interaction between the adapter and the provenance providers so that the \ph{} can provide the provenance service operations to each workflow and
choreography. The controller combines the four methods: record, validate, retrieve and migrate into the realization of the two operations provided by the \ph{}: Collect and Retrieve. 
For the \textit{collect provenance data operation} the controller receives, validates and relays the  provenance information to the providers.
For the \textit{operation retrieve provenance information}, it combines the methods retrieve and validate.
In both cases the validate method is a crucial step (more details in \autoref{ssec:methods-operations}).
The controller has a key storage integrated to be able to verify signatures and identify participants.
If the data is not valid, for instance the signature does not match the signed data or the signer is unknown, the data is rejected and not relayed to the providers for storage.

\subsection{Methods and Operations}
\label{ssec:methods-operations}

As mentioned above, the four main internal methods of the \ph{} are used in different combinations to realize the two external operations (cf. \autoref{fig:arch}).
During the execution and adaptation of workflows and choreographies the \ph{} constantly collects provenance data on a very detailed level, including on per-workflow-activity level.
For all practical purposes, the data collected about experiment execution and applied changes need to be authenticated. This can be done, for instance, by employing public/private key signature algorithms on e.g. input, workflow version, and output data produced by the participating \ac{WfMS} environments.

The \textit{Record} method selects appropriate provider components for a certain workflow type out of the available providers and uses them to store the provenance information.
The information needed for ensuring the provenance of workflow runs/executions is input data, the executed workflow model version and its output data.
The actual provenance information is paired with information about the corresponding choreography instance (typically using a reference to the corresponding choreography instance).
This does not only ease the retrieval but enables the attribution of an execution and data to a certain origin.
For an adaptation on a workflow or a choreography, the provenance information might consist of the description of the actual change performed, the new version of the workflow/choreography and a reference to the preceding one.
Data is validated (with the validation method) before it is actually handed over to a provider for storage.

The \textit{Retrieve} method is used to fetch the desired provenance information from the provider components via their interfaces. The information is identified and retrieved using the corresponding choreography instance ID and/or workflow IDs. The actual data retrieval is done by each provider itself and returned to the \textit{retrieve} method.
After retrieval, the information is validated (with the validation method) before it is handed over to the adapter component, i.e. the \ph{} interface implementation. The validation is used to rule out storing errors or tampering on the data storage and to guarantee the validity of the data and the freshness of the validity check. The retrieved information should then be presented to the user in an implementation specific manner.

During \textit{Validation} the provided signature is verified in the controller component.
Similarly, the signed data is validated and the signee is identified.
When the \textit{Record} method is called, the signature gets verified before the data is \enquote{recorded}.
If the signature can not be verified or the signee not be identified,
the information is rejected and not considered further, hence is not \enquote{recorded}.
The failed signature verification is also communicated to the user.
When calling the \textit{Retrieve} method, the provenance information is fetched from the provenance provider and then validated.
If two or more providers are present in a \ph{}, retrieved data is not only validated but the data of the different providers
and each validation status is compared to one another to identify possible discrepancies.
The status of the validation is also communicated to the user.
Especially if the data storage is at a remote location an adversary might be able to change stored data. If this is done by a participant then he is also able to produce a \enquote{valid} signature of the changed data.
To be able to validate signatures, the participants' keys need to be exchanged beforehand and stored in the \ph{}.

The \textit{Migrate} method is only used if stored information has to be transferred to a new type or instance of storage, in case an addition or change of instances is desired or needed.
It provides the ability to retrieve all stored provenance information from a provider component at once.
Similar to the retrieval of individual objects, all data is validated using the validation method, before it is migrated.

Addition of storage instances means that storage is expanded by a new instance of an already existing technology or a new instance of a not yet used storage technology, e.g. data is stored in an SQL database and now will be copied to a second one or will now also be stored in a flat file.
Change of storage on the other hand means that one storage instance replaces another one.
This can be done within a particular storage technology or done by replacing one technology with another, e.g. data is stored in a flat file and will now be stored in an SQL database, hence it will be migrated.
The employed technology has also implications on the complexity of such a migration because of the difference in features and of their characteristics.
Migrations can be triggered both automatically or manually by an administrator; the actual procedure for migration is out of the scope of this paper as related work like \cite{ART-2014-11} is available.
It is important to note that the cost of the \textit{Migrate} method must be considered, especially when it comes to blockchain technology where the needed information can be spread virtually over the whole ledger.
The migration step might also involve purging data from the source after a successful data transfer.
Here, blockchain technology might also pose additional challenges.

\subsection{Requirements on WFfMSs}
\label{sec:requirements-on-wfms}

The \ph{} architecture introduced above implies several assumptions about the information needed from a \ac{WfMS} and related service middleware so that it can meaningfully collect, process and return provenance information about workflow/choreography changes. From these assumptions we can derive the minimum requirements on \acp{WfMS} and service middleware. 

First, all provenance data produces by the involved \acp{WfMS}, workflow caller and choreography initiator and the information used to identify workflows and choreographies need to be signed and the signature made available to the  \ph{}.
This also implies that the \ac{WfMS} and the service middleware used need to be able to identify the choreography and workflow instances; the concrete technique used for that is technology and implementation specific.

Second, all messages notifying actual changes in workflow/choreography models and/or instances need to be signed too so that both evolution provenance and ad-hoc change provenance can be enabled. 
All private/public key pairs used have to be generated beforehand and made available to the \ph{}.

Third, the key exchange and choreography participant identification has to be done before participants can engage in collaborative scientific workflows.
While the key exchange is trivial as the  key is made public, e.g. alongside the workflow, the participant identification could be done following principles such as trust on first use (TOFU) or trust upon first use (TUFU) and the web of trust. Identification via other channels is also possible and if desired needs to be implemented accordingly.

\section{Design and Implementation}
\label{sec:implementation}

In this section we elaborate on design, technological decisions and implementation details. We discuss several specific aspects regarding our proof-of-concept implementation, available
at: \url{https://github.com/ProvenanceHolder/ProvenanceHolder}

\subsection{Signature algorithm} 

To be able to attribute data to a certain entity as postulated by one of the requirements, we decided to use public/private key signature scheme.
Using this signature scheme not only enables the identification of signers but with the public key meant to be public also not violating privacy or security concerns when storing it on an immutable (public) ledger.
This kind of signature scheme or algorithm makes it possible to identify signers since a signature can
only be validated with a certain key. Key owners are identified before engaging in collaboration and
keys are stored accordingly. Hence, if a signature can be validated, it can be attributed at the same time.
A public/private key algorithm requires the public key to be public and potentially known to everyone in order to validate a signature or encrypt a message (only for the owner of the private key).
Therefore storing the public key on an immutable (public) ledger, like e.g. blockchain, does not impose security or privacy issues. 
While RSA \cite{RSA} is the most popular algorithm in this category,
we chose ed25519 \cite{ed25519} because keys and signatures are significantly smaller without compromising security.
Reference implementations and bindings are also available for a wide range of languages\footnote{\url{https://doc.libsodium.org/bindings_for_other_languages}}, e.g. Python\footnote{\url{https://pynacl.readthedocs.io/en/stable/}} and Java\footnote{\url{https://github.com/terl/lazysodium-java}}. The signature algorithm ed25519 is applied during process execution and when changes are made by the respective parties, i.e. \ac{WfMS} and modeling environment.

\subsection{Controller}

Besides coordinating the individual architecture components (see \autoref{fig:arch}) in order to deliver the two offered operations, the controller has to provide several management features:
\begin{compactitem}[$\bullet$]
    \item Key management: as mentioned above, each and every choreography participant needs to be identified before engaging in collaboration or at least his or her public key has to be saved. As ed25519 keys are \enquote{just} random byte strings only storing them does not suffice for key management and is not efficient when it comes to verifying signatures since there is no information for picking the appropriate key. To the best of our knowledge, there is no ed25519 key management that can do both storing keys and annotating them with the necessary information for automatic and manual key retrieval, therefore we needed to also implement such a key management to suit our needs.
\item Provenance object record: For performance purposes, the controller keeps a record about stored provenance information objects, the \emph{object record}. This way already stored objects can be identified directly without querying any provider, requests for non-existent objects e.g. via \emph{collect} and requests if a certain object exists can also be answered directly. The record of stored objects can also assist with migration procedures between providers.
    \item  Management of linked provenance information objects: The \emph{object record} also keeps track of the linked provenance information objects. In addition to supporting faster search (without the necessity to query the used storage provider component), it also allows for a quick and easy identification of corresponding provenance paths. Each provenance object is recorded by its identifier, i.e. the provenance hash, and contains its predecessor.
\end{compactitem}

\begin{figure}
    \centering
    \includegraphics[width=.2\textwidth]{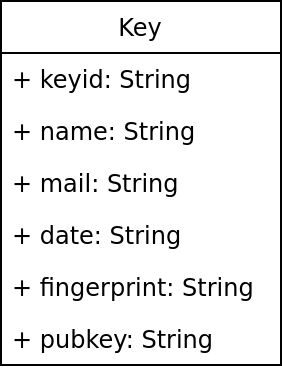}
    \caption{Key object}
    \label{fig:key-object}
\end{figure}

There is information relevant for key management and in particular key retrieval that needs to be stored with the provenance information.
The key identifiers of keys used to sign a provenance information object are stored together with the respective signature.
The to be stored key object (cf. \autoref{fig:key-object}) within the key management has six elements, key id (id),  name (name), e-mail address (mail), creation date (date), fingerprint and the public key (pubkey).
With the fingerprint of the key being a signature with said key over name, mail, date and pubkey, and the id being the last 16 byte of the fingerprint\footnote{This is loosely adapted from OpenPGP (\url{https://datatracker.ietf.org/doc/html/rfc4880}).}.

\begin{figure}
    \centering
    \includegraphics[width=.5\textwidth]{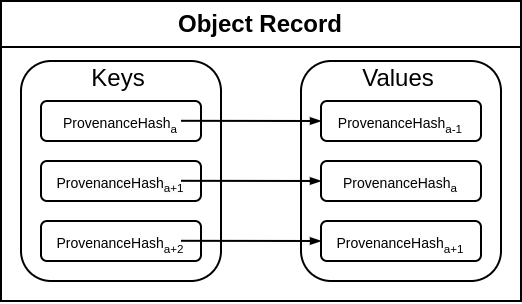}
    \caption{Object record}
    \label{fig:object-record}
\end{figure}

The aforementioned object record (cf. \autoref{fig:object-record}) keeps track of each recorded provenance information object by storing it in a key value data structure with it provenance hash being the key and its predecessor, also identified by its provenance hash, being the value.

\subsection{Adapter}

The main role of the adapter is to provide the two external operations \textit{collect} and \textit{retrieve} (see \autoref{fig:arch}). Furthermore, the adapter is the component used for integrating the \ph{} with any existing \ac{WfMS}. Therefore this is the component that needs to be configured or specifically implemented to allow for integration with the \ac{WfMS} under consideration, the technology it uses and its event models.

In this work we assume that the execution information about workflows and choreographies as well as information about any changes are made available by participating workflow engines to the \ph{} via its adapter. In the current implementation we follow the approach in which it is clearly specified which data is related to change and which to workflow/choreography execution. 
We also assume that the provenance data will be communicated using a  message-oriented middleware, as this mode of communication provides the most advantages for integration of distributed applications and most service middewares support it as well. Besides the information about workflow execution or change the adapter processes the participants signatures. At the moment our implementation is based on these assumptions and realises only the \textit{collect} operation. We are aware that some scenarios may require different integration approaches and may not have explicitly identified change information, for which other solutions need to be investigated in future. Similarly, due to the dependence of the \textit{retrieve} operation on a user-friendly visualization tool and/or specific integration with a WfMS, the retrieve operation's implementation is still under development.

\subsection{Provider}

A provider has the task to store the provenance information.
Most of the members of such a data object can be considered fixed length.
However, when it comes to input, predecessor and possibly also  output,
one has to consider that even a trivial mathematical operation such as the addition has at least two input parameters/operands,
which need to be stored.
As an operation can have an arbitrary number of inputs, even though the length of an input, i.e. the hash of the actual input has a fixed length,
the number of the inputs varies and can be quite big. This fact needs to be accounted for in a suitable way by the technology used. 

\begin{figure}[htp]
    \centering
    \begin{subfigure}[b]{\textwidth}
        \centering
         \includegraphics[width=\textwidth]{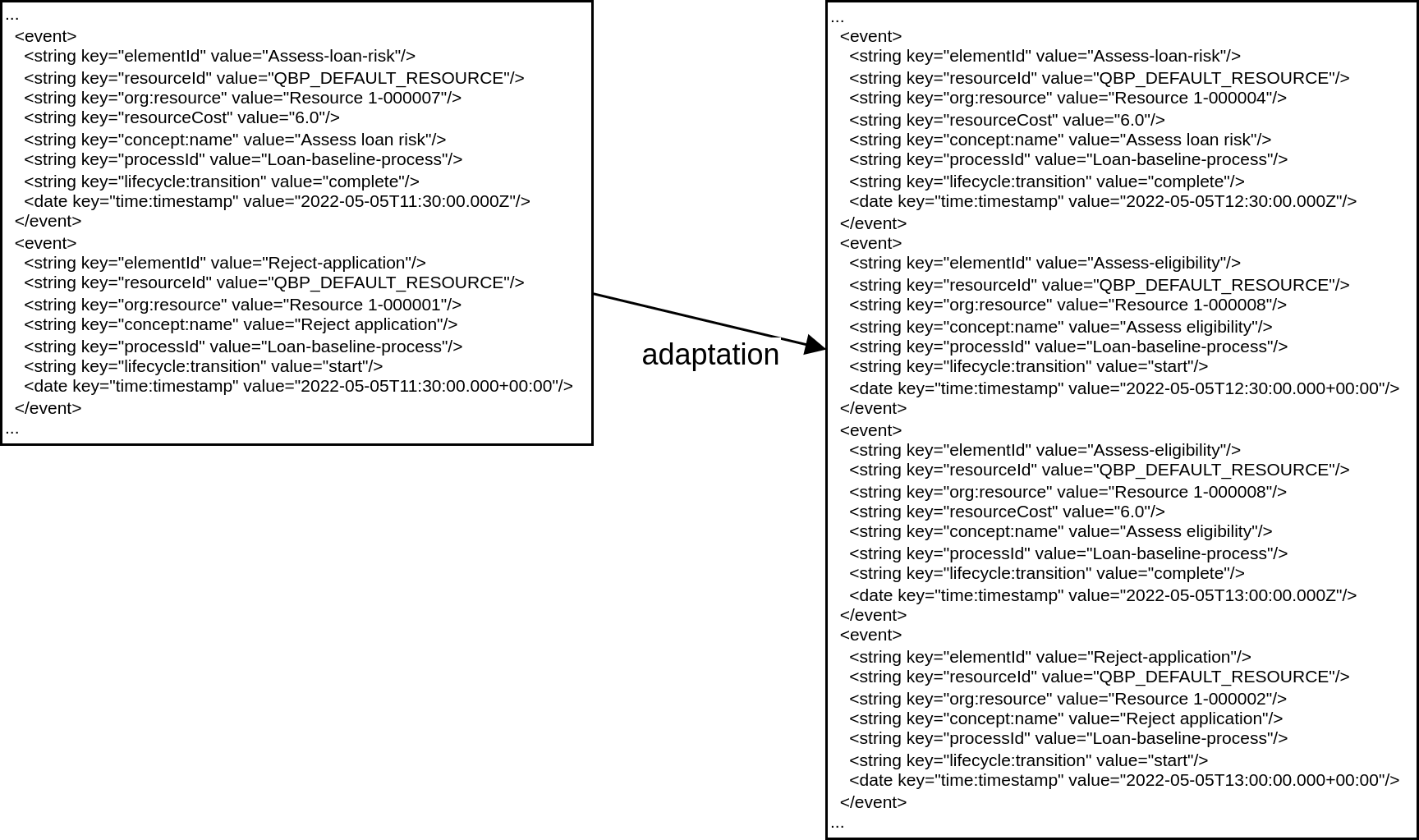}
         \caption{Log events of two workflow instances}
         \label{fig:log-events-to-provenance-info-1}
     \end{subfigure}
     \vfill\vspace{1em}
     \begin{subfigure}[b]{\textwidth}
         \centering
         \includegraphics[width=\textwidth]{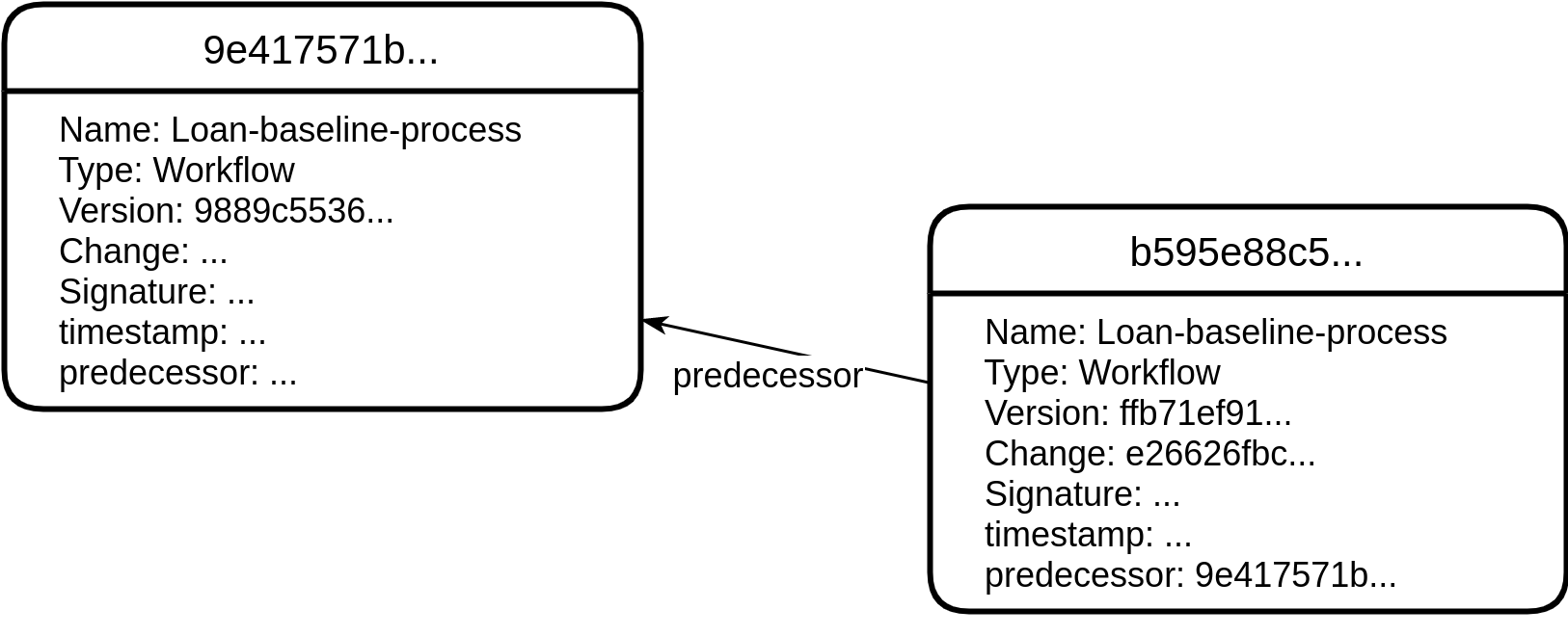}
         \caption{Corresponding Provenance Information objects}
         \label{fig:log-events-to-provenance-info-2}
     \end{subfigure}
    \caption{Log events (XES-compliant) (a) to Provenance Information (recording an adaptation) (b)}
    \label{fig:log-events-to-provenance-info}
\end{figure}

We allow for storing an arbitrary number of predecessors. A predecessor, to a certain provenance information object, is the provenance information object from which the current referenced data was derived from.
This can either be one or more provenance information objects for an execution
or a provenance information object for a change in a model.
In order to be able to store predecessors it is paramount to be able to identify them.
The Choreography instance id and the workflow instance id besides the actual input can be an indication for a predecessor.
However, the actual identification might not be as trivial,
for instance, only because a certain input or output came before in the execution of choreography or workflow does not necessarily mean that it is the right predecessor (cf. \autoref{fig:log-events-to-provenance-info}) or a predecessor at all.
Therefore, the provenance data object structure may need to be extended in our future work on the adapter component, whose task it is to create these objects from the input data.

Changes to a choreography or workflow model, for both instance migration and ad-hoc changes, are stored in the same object type - namely the provenance information object for adaptation.
In case of instance migration the whole workflow/choreography model is captured and in case of ad-hoc changes only the actual change (diff) is recorded.

By storing provenance information in this way we end up with only lists of inputs and outputs/results belonging together in reverse order, which is enough for the provenance of specific data and executions as it is required.
Doubly linked lists are not desired because it should be possible to store the provenance information on append-only data structures such as public ledgers.
A single link suffices since a provenance path can always be back traced to its origin from any element in the path and with the information stored by the controller about recorded provenance information objects it is possible to generate an interlinked list, if necessary.

Consider the following example in \autoref{fig:log-events-to-provenance-info}. In this figure we show an excerpt of an XES\footnote{\url{https://www.xes-standard.org/_media/xes/xesstandarddefinition-2.0.pdf}}-compliant event log of different process instances of a process model from which we can derive and record provenance information.
The logs contain the process related events that would be notified to the service middleware when processes are executed, as well as the events related to changes.
The example log explicitly contains the execution of process instances and implicitly they reflect changes/adaptations to the process model.
As we stated earlier, we record the execution of processes and changes thereof. In our example we focus on capturing the change/adaptation to the underlying process model.
In \autoref{fig:log-events-to-provenance-info-1}, there are two excerpts of an event log which represent an adaption - the first log is the process event sequence/process trace following the original process model, whereas the second event log/trace includes events related to the addition of a new activity to the instance (and the process model).
Execution-wise it is merely a new instance of the process model (instances are recorded but not depicted in the example).
In \autoref{fig:log-events-to-provenance-info-2}, there are two provenance information objects reflecting the change and depicting the predecessor relation between those two objects.
Both objects are identified by their ProvenanceHash, a hash which encompasses all individual elements of the object.

In the following, we will present two proof-of-concept providers, a SimpleStorage provider and a Timestamping provider,
which have different characteristics and serve different purposes in terms of the discussed properties (cf. \autoref{sec:PHproperties}).

\subsubsection{SimpleStorage provider}

Providers can be implemented in any storage technology which suits the respective use case.
We provide a proof-of-concept implementation of a provider which is simple but at the same time provides database capabilities.
The SimpleStorage provider is implemented using SQLite\footnote{\url{https://www.sqlite.org/}} as storage technology, which stores data in files,
provides a feature-rich SQL interface and does not require a full-scale database management system.

This provider addresses the properties P1 and P4 by storing the above mentioned provenance information objects and implicitly the link between them.

The performance characteristics of such a provider need to be evaluated in future research.
\subsubsection{Timestamping provider}

Timestamping in general can be seen as the act of producing a certificate of existence, in our case even without revealing the object which is attested by the certificate.
This can for instance be done for a data object by hashing said object and publishing the produced hash to a medium linear in time, e.g. a public ledger such as blockchain.
By doing so it can be proven that something was known before, which corresponds to the property P2 (cf. \autoref{sec:PHproperties}).
With the Bitcoin blockchain there are several timestamping services such as OpenTimestamps or originstamp \cite{Originstamp2015,Originstamp2018}.
In order to submit hashes to the blockchain, timestamping services accumulate hashes in a Merkle tree and only submit its root hash, or use the root hash as private key for a bitcoin address to which the data is sent.
This is cost efficient since not every hash is submitted individually which saves transaction costs.
At the same time the cost efficiency is traded off by a submission delay of individual hash values.
Since a block containing such a transaction is not instantly created and takes some time until it becomes part of the blockchain anyway, this fact might be negligible.
However, it needs to be evaluated in future work if and when such a delay might prove undesirable.

We employ this technology by submitting the ProvenanceHash, which is a SHA256-Hash, to such a service and keeping track of these submissions. Because of the mentioned delay, the Timestamping  provider implementation has not only to record hashes which are on the blockchain but also hashes pending to become part of the blockchain (as part of a transaction in a block).
The provider has to check periodically for the submission status in order to retrieve the address to which the transaction is ultimately sent to.
The provider has also the task ultimately to make sure that the hash actually becomes part of a block. In order to fulfill this task it might also be necessary to resubmit a hash to the timestamping provider, depending on the respective provider's \ac{SLA}.
By doing so the provider accounts for the eventual consistency property of blockchain technology.
The data to be stored in this case only consists of the ProvenanceHash and the value of the Merkle trees' root of the Merkle tree in which the ProvenanceHash was included.

By timestamping the ProvenanceHash and storing the necessary information to verify said timestamp property P2 is supported.

However, it needs to be investigated further how to assert that one data set existed before or after another, while there are timestamps for each, since time and also the order of blocks is a non-trivial issue within blockchain technology.
It becomes even more complicated if the existing timestamps are on different blockchains.
The authors of \cite{TimeLadleif2020} investigated the aspect time in blockchain-based process execution in their work and also introduce a set of time measures.
These measures will be examined for their usability in the context of timestamping and timestamps in our future research.

\section{Discussion of Open Issues}
\label{sec:discussion}

In this section we discuss open issues not yet covered by our work and highlight directions for future work.

In our previous work \cite{CAiSE18} we identified adaptability (R1), provenance for FAIR (R2), reproducibility for RARE (R3), and trust (R4) as the main requirements of experts on collaborative data processing pipelines and reiterated them in \autoref{sec:PHrequirements}.

Our approach presented here already accounts for all of them:
The evolution of data and models can be retraced through the help of the \ph{} since linked lists of choreography/workflow executions and adaptations are recorded. Thus we address the \textit{provenance} requirement (R2).
The requirement of \textit{reproducibility} (R3) is addressed by recording all executions and changes, which allows for rerunning of the actual data pipeline executions, and thus reproduce them exactly, if, in addition, the necessary access is given to actual data and models and/or changes thereof. In the future we will evaluate if the results can be reproduced in the way experts from different fields require it for different use cases.
The two corresponding requirements, R2 and R3, are addressed by the implemented property P4.

The \textit{trust} requirement (R4) is addressed through digital signature of all recorded data, i.e. by implementing the properties P1 and P2.
Thus, all executions and changes are signed and can be attributed to individual collaboration participants.
While \textit{adaptability} (R1) itself needs to be enabled on \ac{WfMS}-basis, the \ph{}  supports provenance of adaptation since all changes, of the type instance migration or  ad-hoc changes, are recorded in the above mentioned way, too. Hence, by recording all changes and implementing the properties P1, P2 and P4, this requirement is fulfilled.

The adapter component will have the important task of \textit{identifying and selecting the right execution and adaptation data} on the service middleware in a generic manner applicable to different types of \acp{WfMS}.
We currently envision two possible ways of enabling this: a) the needed data is published to specific dedicated topics on the middleware and subscribed by the adapter or b) the adapter picks the appropriate messages and data from the middleware itself.
The first option would pose a rather small integration effort on the adapter, however the requirements on the involved environment, i.e. \acp{WfMS} and modeling environment, are rather high and more intrusive.
The second option, on the other hand, will probably pose no or only minimal additional requirements on the environment but will require higher design and implementation effort on the adapter itself, especially for the cases in which adaptation is not explicitly annotated in the events published on the service middleware; such research is related to the field of process model drift identification or anomaly detection in process execution, as known from the process mining and BPM communities.
The decision of which option to pick and their comparison will be subject of future work as it requires, among all else, rigorous evaluation of quantitative and qualitative performance characteristics.

Besides identifying and selecting the right data, the adapter has also to \textit{identify predecessors} of said data or at least support this identification process. Furthermore, a mapping of data on the middleware with data stored in the \ph{} might be needed.

Furthermore, we find the ability to use existing provenance models, such as PROV-DM to exchange provenance information in a standard manner, beneficial.
While this directly calls for a transformation of the provenance information stored by the \ph{} into this model, being able to import data recorded so that it follows this model might be beneficial as well for the purposes of standardization and reuse. Moreover, there are already available tools for data provenance visualization (e.g. Prov Viewer\footnote{\url{https://github.com/gems-uff/prov-viewer}} \cite{ProvViewer}, ProvViz\footnote{\url{https://github.com/benwerner01/provviz}} \cite{ProvViz}) complying with this standard that could be extended to serve the visualization of the retrieved provenance data.

We already mentioned the topic of zero knowledge proofs, more specifically non-interactive zero knowledge proofs, in the scope of supporting property P3.
Furthermore, in combination with property P4, realised by linking the provenance information objects together, may lead to additional characteristics such as \enquote{chained zero knowledge proofs},
which we will investigate in our future work.

Storage of provenance related artifacts on an immutable (public) ledger is achievable since a) the actual data is not stored there (no security/privacy implication and considerably small cost implication) and b) data from the past is not amended.
While we do not consider this line of research in the scope of our project,  storing all data recorded by the \ph{} on a public ledger might as well be of interest, as it is definitely an unexplored alternative.

\section{Conclusions}
\label{sec:conclusions}

The focus of our work is to enable the trusted provenance and reproducibility of adaptive collaborative data processing pipelines. Our work is based on the workflow management technology for process automation that has proven to bring significant benefits for automating data processing pipelines, and in particular such pipelines that implement in-silico scientific experiments and business analytics pipelines that are in the focus of data-driven, collaborating enterprises. While reproducibility of data processing pipelines has been in the focus of research, the dimensions of collaboration and trust have been abstracted away in available literature, whereas the adaptation of running data processing pipelines has not been considered at all. The work presented in this paper strives towards closing this gap by 1) defining the specific properties of such a service enabling trusted provenance of collaborative and adaptive data processing pipelines, 2) contributing an architecture of a generic provenance service, called \ph{}, 3) a proof of concept implementation   of the approach and 4) identifying the requirements on systems that automate data processing pipelines so that they can integrate with the \ph{} service. 

The  main focus of our future work will be on investigating the applicability of zero knowledge proofs, on the research into the  best alternatives for integration with the \ph{} with focus on the adapter component, on mapping to the provenance information standards available and on visualization of provenance information of change for experts.

%
%
\bibliographystyle{splncs04}
\bibliography{ref}
\end{document}